\documentclass[epsf,usegraphicx]{mn2e}

\title[Red giant pulsations detected from StHA 169]{Red giant pulsations
from the suspected symbiotic star StHA 169 detected in {\kep} data}

\author[]
{Gavin Ramsay$^{1}$, Pasi Hakala$^{2}$, Steve B. Howell$^{3}$ \and\\
$^{1}$Armagh Observatory, College Hill, Armagh, BT61 9DG, UK\\
$^{2}$Finnish Centre for Astronomy with ESO (FINCA), University of Turku,
V\"{a}is\"{a}l\"{a}ntie 20, FI-21500 PIIKKI\"{O}, Finland\\
$^{3}$NASA Ames Research Center, Moffett Field, CA 94095, USA\\
}

\date{Accepted 2014 April 23.  Received 2014 April 23; in original form 2014 February 13}

\begin{document}
\newcommand{\Msun} {$M_{\odot}$}
\newcommand{\Rsun} {$R_{\odot}$}
\newcommand{\kep}{\it Kepler}
\newcommand{\swift}{\it Swift}
\newcommand{\Porb}{P_{\rm orb}}
\newcommand{\nuorb}{\nu_{\rm orb}}
\newcommand{\eplus}{\epsilon_+}
\newcommand{\eminus}{\epsilon_-}
\newcommand{\cd}{{\rm\ c\ d^{-1}}}
\newcommand{\MdotL}{\dot M_{\rm L1}}
\newcommand{\Ldisk}{L_{\rm disk}}
\newcommand{\src}{StHa 169}
\newcommand{\ergscm} {ergs s$^{-1}$ cm$^{-2}$}

\maketitle
\begin{abstract}

We present {\kep} and {\swift} observations of StHa 169 which is
currently classified as a symbiotic binary. The {\kep} light curve
shows quasi periodic behaviour with a mean period of 34 d and an
amplitude of a few percent. Using {\swift} data we find a relatively
strong UV source at the position of StHa 169 but no X-ray counterpart.
Using a simple two component blackbody fit to model the combined
{\swift} and 2MASS spectral energy distribution and an assessment of
the previously published optical spectrum, we find that the source has
a hot ($\sim$10,000K) component and a cooler ($\sim$3700K)
component. The {\kep} light is dominated by the cool component and we
attribute the variability to pulsations in a red giant star. If we
remove this approximate month long modulation from the light curve, we
find no evidence for additional variability in the light curve. The
hotter source is assigned to a late B or early A main sequence star.
We briefly discuss the implications of these findings and conclude
that StHA 169 is a red giant plus main sequence binary.

\end{abstract}


\section{Introduction}

Symbiotic stars are interacting binary systems containing a red giant
star and a hotter component, which can be a white dwarf, a main
sequence star or even a neutron star (see Mikolajewska 2007 for a
review).  A relatively small fraction of these binaries show evidence
for accretion onto the hot component via a disc, while the remainder
show evidence of accretion via the wind from the giant star and, in
some systems, nuclear burning occurs on the surface of the hot
component (see Kenyon \& Webbink 1984). Some systems such as CH Cyg,
have produced jets (e.g. Taylor, Seaquist \& Mattei 1986, Crocker et
al. 2001, Galloway \& Sokolski 2004) and large variations ($\sim$5
mag) in optical brightness over year long timescales (e.g.
Mikolajewski, Mikolajewska \& Khudiakova 1990). More recently,
evidence has been presented which suggests that symbiotic stars could
be progenitors of a fraction of supernovae 1a explosions (e.g. Dilday
et al. 2012).

Given the fact that symbiotic stars contain a red giant star, the
binary orbital periods are the longest (ranging from several hundreds
of days to many years) found in the many different types of
interacting binaries. As such, long-duration photometric surveys such
as OGLE have been used to search for signatures of the binary period
(e.g. Angeloni et al.  2014). In contrast, high time resolution
photometry or spectroscopic observations are required to search for
the presence of phenomena such as accretion whose rate can be
  variable over the medium to long timescale (e.g. Sokolski,
Bildstein \& Ho 2001 and Sokolski \& Kenyon 2003).

The {\kep} misson (Borucki et al. 2010) provides a unique opportunity
to study objects such as symbiotic binaries on short timescales (1
min) and also much longer timescales (the initial {\kep} pointing
lasted approximately 4 years). There were two objects classified as
symbiotic binaries in the {\kep} field -- the previously mentioned
system CH Cyg -- and StHA 169.  This paper presents an analysis of
{\kep} and {\swift} observations of StHA 169.

\begin{figure*}
\begin{center}
\setlength{\unitlength}{1cm}
\begin{picture}(12,12)
\put(-3,6.){\includegraphics{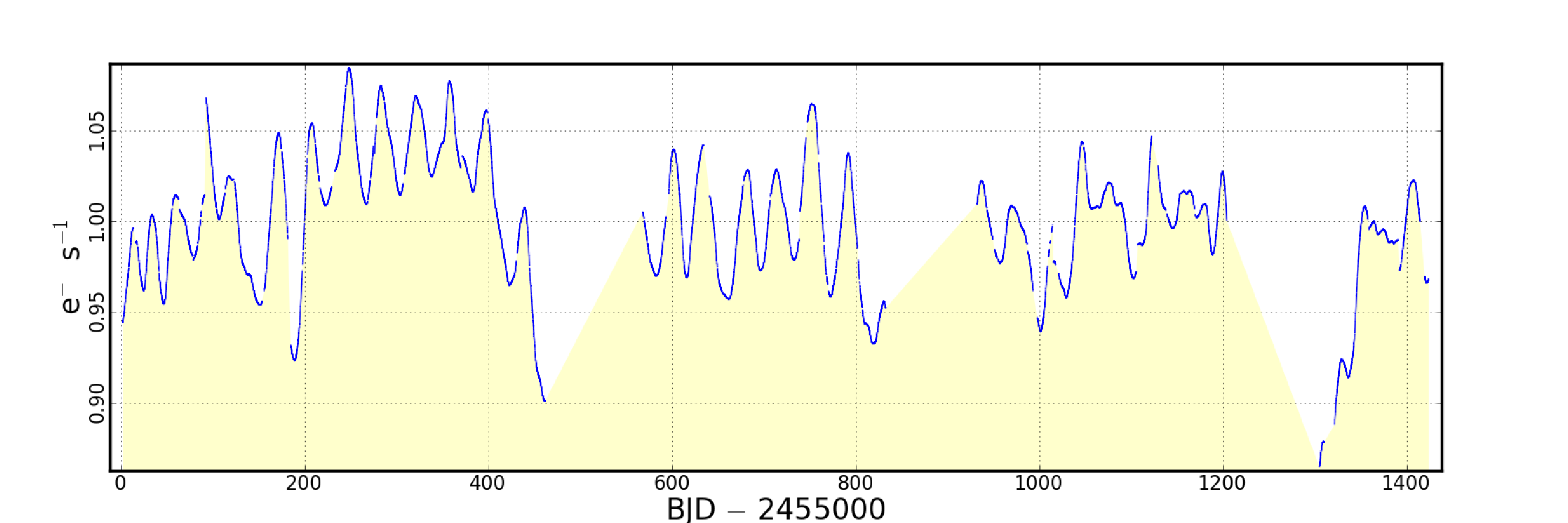}}
\put(-3,-0.4){\includegraphics{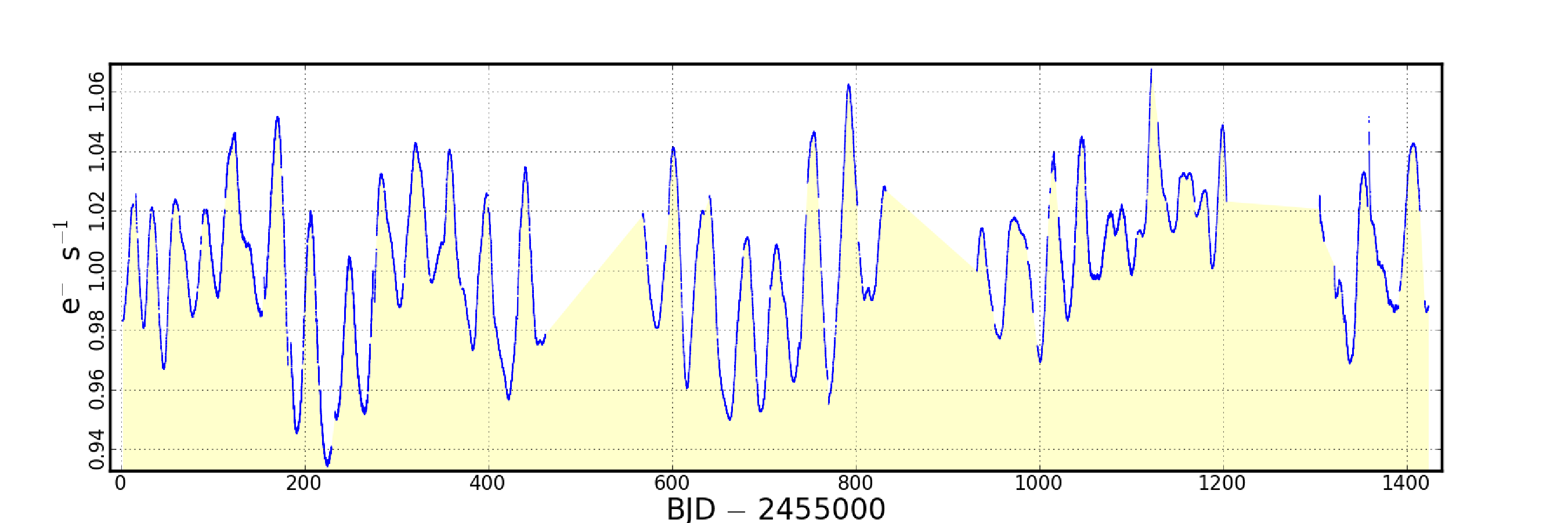}}
\end{picture}
\end{center}
\caption{The upper light curve shows the normalised long cadence data
  of {\src}. The lower light curve has been de-trended and normalised
  and corrected so that there are no step changes in flux between
  different quarters.}
\label{light} 
\end{figure*}

\section{StHa 169}

StHA 169 (also known as NSV 12466 and S169) was classified as a
symbiotic binary by Downes \& Keynes (1988), and is in the catalogue
of symbiotic binaries of Belczy\'{n}ski et al. (2000), as a result of
its optical spectrum which `resembles that of the quiescent phase of
the symbiotic recurrent nova RS Oph'. StHA 169 has the identifier KIC
9603833 in the {\kep} Input Catalogue (Brown et al. 2011) and has a
magnitude of $g$=14.37 and colour $g-r$=1.60 in the {\kep} INT Survey
(Greiss et al. 2012) and $g$=14.12 and $g-r$=1.42 in the {\sl
  RATS-Kepler} Survey (Ramsay et al. 2014). {\src} is recorded in the
ASAS survey of the {\kep} field (Pigulski et al. 2009) as a source
with `no well-defined periodicity in light variations'.

\section{Kepler Photometric Observations}

The detector on board {\kep} is a shutterless photometer using 6 sec
integrations and a 0.5 s readout. There are two modes of observation:
{\it long cadence} (LC), where 270 integrations are summed for an
effective 29.4 min exposure (this includes deadtime), and {\it short
  cadence} (SC), where 9 integrations are summed for an effective 58.8
s exposure.  Gaps in the {\kep} data streams result from, for example,
90$^\circ$ spacecraft rolls every 3 months (called Quarters), and
monthly data downloads using the high gain antenna.

{\kep} data are available in the form of FITS files which are
distributed by the Mikulski Archive for Space Telescope
(MAST)\footnote{http://archive.stsci.edu/kepler}. For LC data each
file contains one observing quarter worth of data whereas for SC data
one file is created per month.  After the raw data are corrected for
bias, shutterless readout smear, and sky background, time series are
extracted using simple aperture photometry (SAP). We note in Table
\ref{kepler-log} the {\kep} Quarters in which LC data was obtained and
also the {\kep} months in which SC mode data was obtained. The first
data to be taken (Q0) started in May 2009 and the final data (Q17)
finished in May 2013.

\begin{table}
\begin{center}
\begin{tabular}{lr}
\hline
Long Cadence & Short Cadence \\
(Quarter)    & (Quarter/Month)\\ 
\\
\hline
0,1,2,3,4,5,6,8,9,10,12,13,14,16,17 & 6/2, 14/2\\
\hline
\end{tabular}
\end{center}
\caption{Journal of {\kep} observations. Each quarter nominally lasts
  3 months with a short gap between months. Short Cadence observations
  are made on a monthly basis. Quarter 17 was truncated to
  approximately 1 month.}
\label{kepler-log}
\end{table}

\subsection{Long Cadence Observations}

Using the data downloaded from MAST we used the `Simple Aperture
Photometry' (SAP) data and removed data which do not conform to
`SAP\_QUALITY=0' (for instance, time intervals of enhanced solar
activity) and then normalised this light curve so that the mean count
rate was unity (Figure \ref{light}).  There are clear flux variations
on a timescale of tens of days and a semi-amplitude of several
percent.

To remove systematic trends in the data (e.g. Kinemuchi et al. 2012)
we used the task {\tt kepcotrend} which is part of the {\tt PyKE}
software (Still \& Barclay
2012){\footnote{http://keplergo.arc.nasa.gov/PyKE.shtml}}. We then
applied a small offset so that there are no discrete jumps in flux
between the different quarters of data. This light curve is also shown
in Figure \ref{light} and shows similar features but with fewer large
flux variations at the start and end of quarters.

\begin{figure}
\begin{center}
\setlength{\unitlength}{1cm}
\begin{picture}(6,6)
\put(-1.7,-0.5){\includegraphics{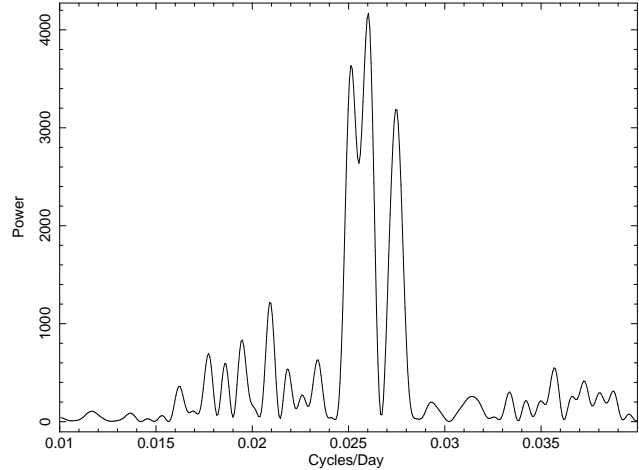}}
\end{picture}
\end{center}
\caption{The Lomb Scargle power spectrum of the light curve
  corrected for systematic trends and jumps between different
  quarters.}
\label{power} 
\end{figure}

We show the Lomb Scargle power spectrum (Lomb 1976, Scargle
1982) of the corrected light curve in Figure \ref{power}. The peaks
correspond to periods of 40.0, 38.5 and 36.3 days. There are also
peaks in the power spectrum at $\sim$200 and 260 days. However, given
the precence of data gaps, the necessary adjustments between each
quarter of data and the known presence of a modulation due to the
{\kep} year (372.5 d) in the data of giant stars (Banyai et al. 2013),
some caution is necessary in interpreting long period signals in power
spectra such as these.

\begin{figure}
\begin{center}
\setlength{\unitlength}{1cm}
\begin{picture}(6,6)
\put(-1.7,0){\includegraphics{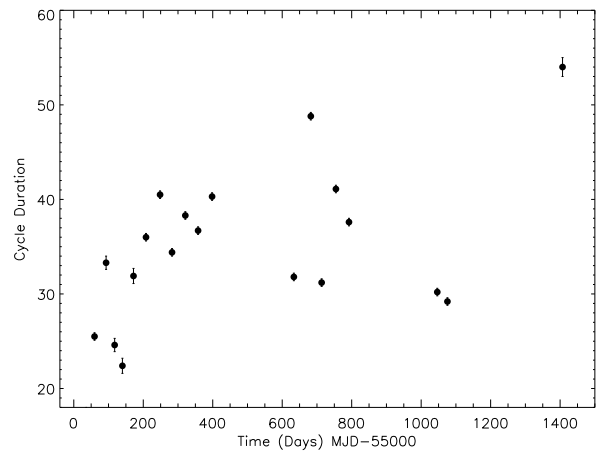}}
\end{picture}
\end{center}
\caption{The length of the 'month' long pulsation period of {\src} as
  a function of time. The length is defined as the time difference
  between a given peak in the light curve and the previous peak.}
\label{length} 
\end{figure}

To further investigate the nature of the light curve, we determined
the time of maximum for every peak in the light curve by eye (the
error on the time of maximum was generally 0.3 days which is very much
smaller than the range of duration of each cycle) and then calculated
the time difference between successive peaks. The duration of each
cycle is shown in Figure \ref{length} and shows a considerable range
in the duration of each cycle, ranging from 23 to 52 days, with a mean
of 34.2 d and $\sigma$=8.4 d.

\subsection{Short Cadence Observations}

{\src} was observed in two quarters (6 and 14) using Short Cadence
mode, which provides photometry with effective exposure times of 58
sec and allows the short term photometric behaviour to be studied in
more detail. To remove the effects of systematic trends and also the
25--50 day pulsation period we used the {\tt PyKE} task {\tt
  kepflatten}. After normalising the data by dividing the light curve
by the mean flux we found the rms was 0.00048 and 0.00049 for data in
quarters/months 6.2 and 14.2 respectively. This result shows very
little evidence of short period variability in the {\kep} light curve.

\section{{\sl Swift} Observations}

{\src} lies $\sim$5 arcmin distant from the $\gamma$-ray burst
GRB060105 and therefore within the field of view of the X-ray (XRT)
and optical/UV (UVOT) instruments on-board {\sl Swift}. Observations
were made between 2006 Jan 5th and 11th. However, only in the dataset
comprising ObsId 00175942000 (Jan 5--6th) were observations made
in the UV filters. We therefore restrict ourselves to this dataset.

\begin{table*}
\begin{center}
\begin{tabular}{lrrrrrrrrrrr}
\hline
Filter & Wavelength & Rate  & Flux            & rms & rms$_{exp}$ & rms/ & sfrac & Points & Bin Size & sfrac\\
       & (\AA)      & (Ct/s) & (\ergscm \AA)  & (Ct/s) & (Ct/s) & rms$_{exp}$ & (sec) & \\
\hline
V   &  5468 &62.9$\pm$0.9 &    1.64$\pm0.02\times10^{-14}$&   0.90 &1.47 &0.61 & 0.014 & 13 & & \\
B   &  4392 &72.3$\pm$1.9 &    1.06$\pm0.03\times10^{-14}$&   1.87 &1.54 &1.21 &0.027 &12 & & \\
U   &  3470 &81.6$\pm$2.8 &    1.32$\pm0.03\times10^{-14}$&   2.84 &1.94 &1.45 &0.035 &12 & & \\
UVW1&  2510 &27.8$\pm$0.9 &    1.17$\pm0.04\times10^{-14}$&   0.91 &0.75 &1.21 &0.033 & 7 & & \\
UVM2&  2250 &12.9$\pm$0.6 &    1.09$\pm0.03\times10^{-14}$&   0.57 &0.38 &1.50 &0.044 &13 & & \\
    &       &11.51        &                              &   1.94 &1.70 &1.14 &0.17  &   & 4 & 0.17\\
    &       &11.54        &                              &   1.09 &0.70 &1.56 &0.09  &   & 60 & 0.09\\
UVW2&  1880 &19.8$\pm$0.6 &    1.18$\pm0.03\times10^{-14}$&   0.61 &0.52 &1.27 &0.031 & 9 & & \\
\hline
\end{tabular}
\end{center}
\caption{Details of the Swift UVOT observations made in 2006 Jan 5--6
  where we indicate: the filter and its central wavelength; the mean
  count rate and flux; the rms derived from the given number of
  points. We also indicate the expected rms$_{exp}$ which is
  determined from the mean count rate and sfrac = rms/Rate. We also
  determine sfrac for two different bin sizes and were derived from
  the event data.}
\label{swift-obs}
\end{table*}

We show in Table \ref{swift-obs} the filters in which observations
were obtained. We used the {\tt ftool} task {\tt
  uvotsource}\footnote{http://heasarc.gsfc.nasa.gov/lheasoft/ftools/headas/uvotsource.html}
to determine the mean count rate for {\src} and the corresponding
flux. The exposure time of each UVOT image is generally 500--700 sec
in duration and there were 7 images in the UVW1 filter and 13 images
in the UVM2 filter (other filters had values between these
numbers). Based on the count rates derived from these images we
determined the rms and expected rms (assuming Poisson statistics) and
the corresponding value sfrac (=rms/count rate). These numbers are
given in Table \ref{swift-obs}. In no case was the rms variation
greater than twice the expected rms. We also did the same analysis
using event rate data which was obtained using the UVM2 filter. We
binned the data using different binsizes and show the results for a 4
sec and a 60 sec binsize in Table \ref{swift-obs}. Again the
rms/rms$_{exp}$ ratio is less than 2 (and was for all attempted
binsizes). This finding of very low short term variability is
consistent with that found in the Short Cadence {\kep} observations.

In their study of symbiotic binaries using {\swift} Luna et al. (2013)
found that in the UV 33 non-saturated sources shows rms/rms$_{exp}>$2.
All the sources which did not show significant rms variability were
fainter in that band compared to {\src}. {\src} therefore displays an
unusually low degree of variability compared to the majority of
symbiotic binaries.

In the dataset comprising ObsID 00175942000, the combined X-ray
exposure was 57.8 ksec. However, there is no detection of a source at
the position of {\src}. The count rate for the location of {\src} is
0.000011$\pm$0.000009 ct/s where we have used a source aperture of 20
arcsec (which excludes a nearby X-ray source) and subtracting a scaled
and much larger background aperture. The XRT is sensitive to a flux of
$\sim1.5\times10^{-14}$ \ergscm for this observation.

\section{Spectral Energy Distribution}
\label{sed}

With the available data of {\src} ranging from the near-UV and
  optical ({\sl Swift}), the near-IR (2MASS) and far-IR (WISE) we can
  obtain the broad spectral energy distribution of {\src} and from
  that assess which binary component is dominant in the {\sl Swift}
  and {\kep} band-passes. However, as symbiotic stars are known to
  show flux variations over the long term some assessment needs to be
  made as to the long term variability of {\src}. Perhaps the most
  spectacular long term variability seen in a symbiotic star is CH Cyg
  which has shown variations of 6 mag in the U band, although the
  variation is much reduced ($<1$ mag) at infrared bands (eg Munari et
  al. 1996). On the other hand, other symbiotic systems show much lower
  levels of long term variability. For instance, in a study of
  symbiotic binaries in the LMC (Angeloni et al 2014), two systems
  showed a variation less than 14\% in the I band over a time span of
  more than 800 d.

However, perhaps the best evidence of long term variability is the
{\kep} data of {\src} itself which spans nearly four years. This shows
a standard deviation of 3.6 percent in the unnormalised light curve
and 2.7 percent in the unnormalised detrended light curve. Other data,
for instance, the g mag from the {\sl KIS} and {\sl RATS-Kepler}
surveys differ by 0.25 mag, while the $V$ mag from the {\sl Swift}
mission and the {\sl Kepler} UBV survey (Everett, Howell \& Kinemuchi
2012) differ by 0.28 mag. {\sl ASAS} had (incomplete) coverage of
{\src} over a 600 day interval and found a standard deviation of 0.13
mag in the light curve (Pigulski et al 2009).

\begin{figure}
\begin{center}
\setlength{\unitlength}{1cm}
\begin{picture}(6,6)
\put(-2.6,-6.5){\includegraphics{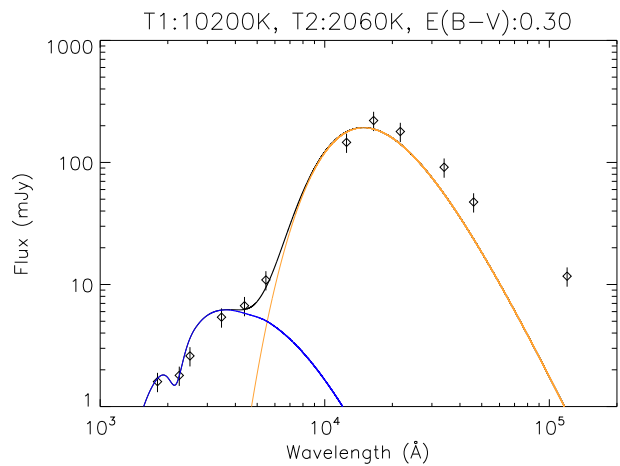}}
\end{picture}
\end{center}
\caption{The spectral energy distribution of {\src} from the near-UV
  to near IR. The absorbed flux unit is milli-Jy and we show the joint
  fit (solid line) from two blackbody models with temperature $\sim$2100K
  and $\sim$10200K and absorption $E_{B-V}$=0.26. We have not used the WISE
  points (the three right most points) in the fit.}
\label{sedfig} 
\end{figure}

We therefore use {\sl Swift} data from the dataset ObsId 00175942000
(as we used in the previous section) since photometric data using six
UVOT filters were made within a short (2 days) time interval. We have
also extracted from the archives the 2MASS and WISE flux measurements.
In Table \ref{flux} and Figure \ref{sedfig} we show the observed fluxes
from these instruments in mJy.

Although cool stars are poorly fit by blackbody models (which we
discuss in the next section) they are simple to use and allow us to
easily determine which stellar component is dominant in the {\kep} and
{\swift} band passes.  The fractional residuals to a two blackbody
model were measured to have a scatter of 18 percent about the fit
(Figure \ref{sedfig}). To account for any changes in the brightness at
different epochs, we therefore assigned the error in each photometric
band to be 18 percent and carried out a set of 200 Monte Carlo
simulations to derive the errors for the parameters. We find T1=10200
$\pm$ 1100K, T2=2060 $\pm$ 140K, The ratio of projected areas (red
giant/blue star) is 29800 $\pm$ 12000 and E$_{B-V}$ = 0.30 $\pm$
0.10. Although the WISE flux measurements are shown in Figure
\ref{sedfig}, we have not actually used them in the fit since emission
from dust (originating in the immediate vicinity of {\src}) is likely
to be strong at these wavelengths.  Indeed, given that the WISE flux
measurements are in excess of the model fit, this points to the
presence of dust in the {\src} system.

What this simple modelling shows is that the {\kep} band-pass
(4200--9000\AA) is dominated by the flux from a cool star, while the
UV and blue flux in the {\sl Swift} UVOT is dominated by a hotter
component. In the $V$ band the cool component contributes 70 percent
of the observed flux.

\section{The nature and distance of the binary components}

Although the above modelling gives approximate temperatures for
  the two components, it is well known that cool stars (of any
  spectral class) are very poorly fit by a blackbody model. To obtain
  a better estimate of the temperature of the cool star in {\src} we
  used the spectral atlas of Pickles (1998) and the published spectrum
  of {\src} (Fig. 4 of Downes \& Keynes 1988). Given that the {\kep}
  data strongly supports the view that the cool star is a pulsating
  giant, we compared red giant spectra given in Pickles (1998) with
  {\src}. Based on the general shape of the spectra of {\src} and the
  dip near 5452 \AA, we assign a spectral type of M2III--M3III to
  {\src}. Assuming this range of spectral type we can then assign
    a temperature of 3650--3750 K (Straizys \& Kuriliene 1981) which
    is considerably hotter than the fit derived using two blackbody
    components (2100 K) but physically more realistic for a pulsating
    red giant star.

Using the information to hand we can make estimates on the
  distance to {\src} using different assumptions and assess the nature
  of the hot component. For an observed $(J-K)$=1.17 and
  $(J-K)_{o}$=0.90, Tabur et al. (2009) indicate that this colour
  implies $M_{K}\sim-4.5$ and a resulting distance of $\sim$4.5kpc,
  whilst the same authors find that the absolute magnitude of the tip
  of the red giant branch is $M_{K}$=--6.85 implying a distance of
  13.7 kpc if {\src} was on at this stage in its evolution. We can
  also make an approximate estimate of the distance to the system by
  taking the observed relationship between the pulsation period of red
  giants and $M_{K}$ (Tabur et al. 2010). For a pulsation period of 34
  d we estimate $M_{K}\sim-5.8\pm0.5$ and a distance of 8.6$\pm$2.4
  kpc.

Although there is a significant degree of uncertainty (and range)
  in these distances (and it is possible the $K$ band may have
  residual dust contamination, see \S \ref{sed}), we conclude that
  {\src} is at least 4 kpc distant. {\sl Gaia} should be able to
determine the distance to {\src} with an accuracy of $\sim$20 percent.

Our blackbody fits (\S \ref{sed}) suggest that the temperature of the
hotter component is $\sim$10000K which is very similar to Vega (an AOV
star). Our blackbody fits indicate that the hotter component
contributes 30 percent of the overall flux in the $V$ band, and that
an AOV star would lie at a distance of 5.3kpc. On the other hand if
the hotter component was a pure hydrogen white dwarf with a
temperature of 10000 K, it would have an absolute magnitude of
$M_{V}$=12.1 (Bergeron, Wesemael, \& Beauchamp 1995), implying a
distance of 26 pc. Even if there was an accretion disk around a white
dwarf (which could brighten the absolute magnitude by 2 mag), the
distance would be very much closer than that inferred using the cooler
component.  We conclude that the hot component in {\src} is not a
white dwarf. However, the hotter star must be less massive than the
red giant star (estimated to be 3.3--3.5 \Msun). An AO V star has a
mass of 2.2 \Msun while a B8 V star has a mass of 3.0 \Msun (Straizys
\& Kuriliene 1981).

\begin{table}
\begin{center}
\begin{tabular}{rlr}
\hline
Wavelength & Filter & Flux \\ 
(\AA)      &        & mJy \\
\hline
1880 & Swift UVW2 & 1.6$\pm$0.1 \\
2250 & Swift UVM2 & 1.8$\pm$0.1 \\
2510 & Swift UVW1 & 2.6$\pm$0.1 \\
3470 & Swift U    & 5.4$\pm$0.2\\
4392 & Swift B    & 6.7$\pm$0.2\\
5468 & Swift V    & 16.0$\pm$0.2\\
12500 & 2MASS J   & 146.0$\pm$3.1\\
16500 & 2MASS H   & 220.5$\pm$3.2\\
21700 & 2MASS K   & 179.1$\pm$2.6\\
34000 & WISE W1 & 91.4$\pm$2.0\\
46000 & WISE W2 & 47.4$\pm$0.9\\
120000  & WISE W3 & 11.7$\pm$0.3 \\
\hline
\end{tabular}
\end{center}
\caption{The observed flux in mJy of {\src}. The errors for {\sl
    Swift} fluxes come from the standard deviation of the observed
  fluxes in {\sl Swift} data obtain during 2006 Jan 5--6.  The error
  on the 2MASS and WISE fluxes come from the error on the magnitude in
  the 2MASS and WISE catalogues.}
\label{flux}
\end{table}

\section{Discussion and Conclusions}

The {\sl Swift}-2MASS spectral energy distribution of {\src}
  indicates there are two components, one relatively hot and one
  relatively cool. Using additional information such as the existing
  optical spectrum of Downes \& Keyes (1988) our best estimate for the
  temperature of the two components is 10000K and 3700K. The {\kep}
observations therefore sample the cooler (and physically much larger)
component, while the {\swift} UV and blue filters sample the hotter
component. The {\kep} light curve shows quasi periodic behaviour
  with a mean period of 34 d. Given that this period is not stable,
it is clearly not the signature of a binary orbital period. Rather it
indicates that the cool component is a pulsating red giant star. This
is consistent with the suggestion made by Downes \& Keyes (1988)
that the cool component in {\src} has a M spectral type and resembles
the red giant in the recurrent nova RS Oph (P$_{orb}$=460 d; Dobrzycka
\& Kenyon 1994; M0--M2 III, Dobrzycka et al. 1996).

Further confirmation of the evolutionary status of the red giant in
StHA 169 is provided through its frequency spectrum. {\kep} photometry
has been used extensively to characterize red giants as to their
membership on the RGB or the AGB (Chaplin et al. 2013). The {\kep}
light curve of {\src} is very similar in character to, say, the red
giant KIC 2986893 (B{\'a}nyai et al. 2013) which has a mean period of
21.7$\pm$=2.2 d. However, given that {\src} has a wide range of
  pulsation period, it is possible that it is a Semi Regular Variable
  (cf Soszy\'{n}ski, Wood \& Udalski 2013). B{\'a}nyai et al. (2013)
showed that M giants separate into three distinct groups according to
their period structure. StHA 169 and KIC 2986893 belong to group 1 --
red giants with periods between 10--100 days. Group 1 stars with a
period similar to that found in {\src} tend to lie the upper Red Giant
Branch (see Kiss \& Bedding 2003, 2004) and are pulsating due to first
and second overtone modes. (Given the main sequence lifetime of a
  3.3 \Msun star is 500 Myr, the system is at least this old).

The identification of {\src} as a symbiotic star lies solely with
  the optical spectrum presented in Downes \& Keyes (1988). The
  spectral energy distribution as derived from {\swift} and 2MASS
  photometry and presented in \S \ref{sed} is consistent with {\src}
  being a binary system. Similarly, our detection of pulsations in the
  {\kep} data clearly demonstrates that the cool star is a red
  giant. However, determining the nature of the hot component in
  symbiotic stars is not a trivial task (see, for instance, Sokoloski
  \& Bildsten 2010 who recount the quarter of a century debate on the
  nature of the hot star in the Mira AB system). Our spectral energy
  distribution shows that an isolated white dwarf or a white dwarf
  with an accretion disk would not lie at the same infered distances
  for the red giant component. Instead our results favour that the hot
  star is more likely to be a late B or early A main sequence
  star. The absence of short period variability in the UV and the non
  detection in X-rays suggest that accretion was not taking place at
  the time of these {\swift} observations.
 
There are at least two other sources which bear some similarity to
{\src}: XX Oph and AS 325 which are thought to consist of a Be star
and a red giant secondary (Howell, Johnson \& Adamson 2009). Indeed,
AS 325 was originally taken to be a symbiotic system.  The fact that
the optical spectrum of {\src} (Downes \& Keyes 1988) shows the Balmer
lines (and He II 4686\AA) in emission may indicate that the B/A star
is an emission star (either through a wind or accretion). Stars like
these are interesting from a binary evolution point of view.
Determining the binary orbital period is a key step but will be
difficult to disentangle the signature of the binary period from the
red giant pulsations.

\section{Acknowledgments}

{\kep} was selected as the 10th mission of the Discovery Program.
Funding for this mission is provided by NASA, Science Mission
Directorate.  The {\kep} data presented in this paper were obtained
from the Multimission Archive at the Space Telescope Science Institute
(MAST). STScI is operated by the Association of Universities for
Research in Astronomy, Inc., under NASA contract NAS5 26555. Support
for MAST for non HST data is provided by the NASA Office of Space
Science via grant NAG5 7584 and by other grants and contracts. This
work made use of PyKE, a software package for the reduction and
analysis of Kepler data. This open source software project is
developed and distributed by the NASA Kepler Guest Observer
Office. Armagh Observatory is supported by the Northern Ireland
Government through the Dept of Culture, Arts and Leisure. We thank the
anonymous referee for a useful report.

\end{document}